\documentclass[a4]{epl2}
\usepackage{graphicx}
\usepackage{epsfig}
\usepackage{color}

\title{\textcolor[rgb]{0.00,0.00,0.00}{On relativistic   harmonic oscillator}}
\shorttitle{\textcolor[rgb]{0.00,0.59,0.00}{On relativistic  harmonic oscillator}}

\author{A. I. Arbab\inst{}\footnote{arbab.ibrahim@gmail.com}}

\institute{
  \inst{} {Department of Physics,
Faculty of Science, Qassim University,  Buraidah 51452, KSA}}


\abstract{A  relativistic quantum harmonic oscillator in 3+1 dimensions is derived from a quaternionic non-relativistic quantum harmonic oscillator. This quaternionic equation also yields the Klein-Gordon wave equation with a covariant (space-time dependent) mass. This mass is quantized and is given by $m_{*n}^2=m_\omega^2\left(n_r^2-1-\beta\,\left(n+1\right)\right)\,,$ where $m_\omega=\frac{\hbar\omega}{c^2}\,,$ $\beta=\frac{2mc^2}{\hbar\,\omega}\, $, $n$, is  the oscillator index,  and $n_r$ is the refractive index in which the oscillator travels. The harmonic oscillator in 3+1 dimensions is found to have a total energy of $E_{*n}=(n+1)\,\hbar\,\omega$, where $\omega$ is the oscillator frequency. A Lorentz invariant solution for the oscillator is also obtained. The time coordinate is found to contribute a term $-\frac{1}{2}\,\hbar\,\omega$ to the total energy. The squared interval of a massive oscillator (wave) depends on the medium in which it travels. Massless oscillators have null light cone. The interval of a quantum oscillator is found to be determined by the equation, $c^2t^2-r^2=\lambda^2_c(1-n_r^2)$, where $\lambda_c$ is the Compton wavelength. The space-time inside a medium appears to be curved for a massive  wave (field) propagating in it.}
\begin{document}
\maketitle
\baselineskip=20pt

\section{\textcolor[rgb]{0.00,0.07,1.00}{Introduction}}

Shr\"odinger equation is very successful in identifying the physics of a quantum simple harmonic oscillator. Simple harmonic motion is of prime importance in physical world. It finds a lot of applications in physics. For instance, atoms in a crystal lattice execute simple harmonic motion. While a classical harmonic motion is determined by Newton's laws, a quantum analogue reveals the quantization of the oscillator energy that was left unrestricted  in the classical regime. 

While Shr\"odinger equation can describe a non-relativistic harmonic oscillator, a relativistic one is left to be described by a relativistic quantum equation. This possibility may allow us to envisage an alternative method to quantize the field of a relativistic quantum boson particle. Note that the electromagnetic field has been quantized as consisting of a collection of an infinitely large number of  harmonic oscillators. A relativistic oscillator picture can help quantize a massive field. It can be applicable to any confined system where relativistic quantum effects become dominant, e.g., nuclear and quark model of matter.

The total energy of the oscillator reflects the degree of freedom the oscillator has. A harmonic oscillator in 3 - dimensions is described by its total energy, $E_n=(n+\frac{3}{2})\, \hbar\,\omega$\, where $n=0, 1, 2, \cdots $, and $\omega$ is the oscillator frequency \textcolor[rgb]{0.00,0.07,1.00}{\cite{dirac-klein}}. In quantum mechanics, we deal with probability in identifying the position and momentum of a particle. For instance, the position of the oscillator is not well determined  a priori. Owing to the theory of relativity, a particle in space-time follows a path, world- line, that is defined by the squared interval, $s^2=c^2t^2-r^2$. Massless particles move along a null world line, $s^2=0$, with velocity equals to that of light in vacuum, while massive particles are describe by a time-like interval, $s^2>0$, and so their velocity is less than that of light in vacuum. However, when the medium is not free, such an interval will be altered and depends on the refractive index of the medium present. This situation could be present when we think of light propagation in curved space.

The harmonic oscillator model is very successful in describing the crystal lattice of a solid that gives a correct thermodynamic description pertaining to specific heat capacity of the solid. However, a field theoretic model is  based on treating the field as an infinitely large number of harmonic oscillators. It is thus instructive to consider an extension of a non-relativistic harmonic oscillator to a quantum relativistic one where relativistic effect are blended with quantum one. Therefore, one may suggest the use of a relativistic quantum oscillator in describing the gravitational field, when the effect of curvature and quantum mechanic become imminent.

The harmonic oscillator studied by Shr\"odinger is non relativistic and is thus valid only under certain conditions. The search for a relativistic quantum harmonic oscillator can be traced back to several authors \textcolor[rgb]{0.00,0.07,1.00}{\cite{relativ, relativ0, relativ1, relativ2, relativ3, relativ4}}. It is argued by Feynman that a covariant harmonic oscillator can explain the quark model \textcolor[rgb]{0.00,0.07,1.00}{\cite{relativ3}}. We pose the question that how the path of a quantum particle is described by virtue of special relativity and quantum mechanics. To answer these questions, we investigate the evolution a harmonic oscillator in 3+1 dimensions (space-time continuum).

To this aim we introduce the quaternionic Shr\"odinger equation for the harmonic oscillator. In quantum field theory, the field is modeled by a collection of an infinitely large number of harmonic oscillators. In such a case a relativistic picture of the oscillator becomes inevitable. While Shr\"odinger equation is not a relativistic wave equation, the quaternionic Shr\"odinger equation, however, yields a relativistic quantum harmonic oscillator equation that is analogous to the Klein-Gordon wave equation.  This urges us to employ quaternions to describe physical always where we expect to obtain un-expected interesting results that may have been connected with another physical phenomena. This may result from the structure of the space-time in which quaternionic quantities act. It is interesting to see that a non-relativistic quaternion equation led steadily to a relativistic one. Hence,  a path from one paradigm to another is thus opened.

Recall that the relativistic and quantum nature of the  oscillators are not independent, they are constrained by the quantum relativistic interval, \emph{viz.}, $r^2-c^2t^2=\lambda_c^2(n_r^2-1)$,  where $\lambda_c=\frac{\hbar}{mc}$ is the Compton wavelength of the oscillator, and $n_r$ is the refractive index of the medium in which the oscillator propagates.  The total energy of the oscillator in 3+1 dimension is found to be $E_{*n}=(n+1)\,\hbar\,\omega$, where the contribution of the time coordinate in the Hamiltonian is equal to $-\frac{1}{2}\hbar\omega$.

This may suggest that the time acts like an energy sink. Or equivalently, this term may be considered as a relativistic correction to the 3 - dimensional harmonic oscillator. In d+1 dimensions, the total energy is $E^d_{*n}=(n+\frac{d-1}{2})\,\hbar\,\omega$. Interestingly, the ground state energy of  1+1 dimensional relativistic quantum harmonic oscillator is zero.

\section{\textcolor[rgb]{0.00,0.07,1.00}{Quaternionic Shr\"odinger equation}}

A classical 3-dimensional harmonic oscillator is described by the Hamiltonian
\begin{equation}
H=\frac{p^2}{2m}+\frac{1}{2}\,m\omega^2r^2\,,
\end{equation}
where $m$ and $\omega$ are the mass and frequency of the oscillator.
In quantum mechanics, it is described by the Hamiltonian
\begin{equation}
H=\frac{P^2}{2m}+\frac{1}{2}\,m\omega^2X^2\,,
\end{equation}
where $X$ and $P$ are now hermitian operators, and expressed mathematically as $P=-i\hbar\vec{\nabla}$ and $X=r$. The state of the oscillator is described by Shr\"odinger equation
\begin{equation}
H\,\psi(r, t)=E\,\psi(r, t)\,,
\end{equation}
where $E$ is the total energy of the oscillator.

Let us now extend $\psi$, $P$ and $X$ to become quaternions, as follows
\begin{equation}
\tilde{\Psi}=\left(\frac{i}{c}\,\psi_0\,, \vec{\psi}\right)\,,\qquad \tilde{P}=\left(\frac{i}{c}\,E\,, \vec{p}\right)\,, \qquad \tilde{X}=(ic t\,, \vec{r})\,.
\end{equation}
Applying Eq.(4) in Eq.(3), using Eq.(2), yields the quaternionic (extended) Shr\"odinger equation for the harmonic oscillator, \emph{viz.,}
\begin{equation}
\frac{1}{2m}\tilde{P}^*\tilde{P}\,\tilde{\Psi}+\frac{1}{2}m\omega^2\tilde{X}^*\tilde{X}\,\tilde{\Psi}=E\tilde{\Psi}
\end{equation}
Upon using Eq.(4) and the product rule for two quaternions, $\tilde{A}=(a_0,\vec{a})$ and $\tilde{B}=(b_0, \vec{b})\,,$ \emph{i.e.,}
$
\tilde{A}\,\tilde{B}=(a_0b_0-\vec{a}\cdot\vec{b}\,\,, a_0\vec{b}+\vec{a}\,b_0+\vec{a}\times\vec{b}),$ one finds
\begin{equation}
\frac{1}{c^2}\frac{\partial^2\psi_0}{\partial t^2}-\nabla^2\psi_0+\frac{2m}{\hbar^2}\left( \frac{1}{2}\,m\omega^2(r^2-c^2t^2)-E\right)\,\psi_0=0\,,
\end{equation}
after equating the scalar parts of the resulting equations. The vector part, $\vec{\psi}$, also  satisfies the same differential equation in Eq.(6), \emph{viz}.,
\begin{equation}
\frac{1}{c^2}\frac{\partial^2\vec{\psi}}{\partial t^2}-\nabla^2\vec{\psi}+\frac{2m}{\hbar^2}\left( \frac{1}{2}\,m\omega^2(r^2-c^2t^2)-E\right)\,\vec{\psi}=0\,,
\end{equation}
Owing to this vector property, $\psi_0$ and $\vec{\psi}$ can be availed  to study the photon dynamics and the related electromagnetic properties.
Equations (6) and (7) are symmetric equations in space and time, and thus are Lorentz invariant. It is thus invariant under Lorentz transformation. In a covariant form, Eq.(6) can be expressed as
$$
\left(\partial_\mu\partial^\mu-\frac{2m}{\hbar^2}\left(\frac{1}{2}\,m\omega^2\,x_\mu x^\mu+E\right)\right)\psi_0=0\,.
$$
Equation (7) can be seen as a Klein-Gordon wave equation with a space-time - dependent mass "effective mass" as
\begin{equation}
m_*^2=m^2\left(\frac{\omega^2}{c^2}\,(r^2-c^2t^2)-\frac{2E}{mc^2}\right)\,.
\end{equation}
It is evident from Eq.(8) that $m_*$ is imaginary for $r^2-c^2t^2<0$ (time-like interval), and is zero for\, $r^2-c^2t^2=\frac{2E}{m\omega^2}$\,. This equation is that due to a hyperbola. Therefore, an oscillator moving along this hyperbola is massless.
Equation (6) suggests that a time-independent (stationary) state along the null path with  $r=\pm\,ct$ is described by
\begin{equation}
\nabla^2\psi_0+\frac{2mE}{\hbar^2}\,\psi_0=0\,.
\end{equation}
The solution of Eq.(9) is that of a standing wave.
With some scrutiny, Eq.(6) can be seen as  relativistic - Shr\"odinger equation for the harmonic oscillator with a space-time dependent potential energy that is given by
\begin{equation}
V(r, t)=\frac{1}{2}\,m\omega^2(r^2-c^2t^2)\,.
\end{equation}
It is interesting to notice that Eq.(6) is Lorentz invariant, and can be viewed as representing a 4-dimensional harmonic oscillator. The potential $V(r, t)$ is negative for time-like and positive for space-like intervals.
Notice that when $m=0$, Eq.(8) yields, $m_*=0$ too, at any position and at any time.

We now separate the spatial and temporal components of  Eq.(6) as
\begin{equation}
\frac{1}{c^2}\frac{\partial^2\psi_0}{\partial t^2}-\frac{m^2\omega^2c^2}{\hbar^2}\,t^2\,\psi_0=\nabla^2\psi_0+\frac{2m}{\hbar^2}\,(E-\frac{1}{2}\,m\omega^2r^2)\,\psi_0\,.
\end{equation}
Let us write $\psi_0(r,t)=\phi(r)\,\chi(t)$. Applying this in Eq.(11), and equate each part of the resulting equation to $-\lambda$, yield the two equations
\begin{equation}
\frac{1}{c^2}\frac{d^2\chi}{dt^2}-\frac{m^2c^2\omega^2}{\hbar^2}\,t^2\chi+\lambda\,\chi=0\,,
\end{equation}
and
\begin{equation}
\nabla^2\phi+\frac{2m}{\hbar^2}\left(E+\frac{\hbar^2\lambda}{2m}-\frac{1}{2}\,m\,\omega^2r^2\right)\phi=0\,,
\end{equation}
where $\lambda$ is constant. Equation (13) is the ordinary equation of the 3-dimensional harmonic oscillator but with energy $E_*=E+\frac{\hbar^2\lambda}{2m}$. The solution of Eq.(13) in 1 - dimension (x-axis) is given by the normalized eigenfunction \textcolor[rgb]{0.00,0.07,1.00}{\cite{dirac-klein}}
\begin{equation}
\phi_n(x)=\left(\frac{\sqrt{\frac{m\omega}{\pi\hbar}}}{2^nn!}\right)^{1/2}H_n\left(\sqrt{\frac{m\omega}{\hbar}}\,x\right)\,e^{-\frac{m\omega}{2\hbar}x^2}\,,
\end{equation}
where $n=0, 1, 2, \cdots$, having the energy eigenvalues,   $$E_*\equiv E_n=(n+\frac{1}{2})\,\hbar\,\omega+\frac{\hbar^2\lambda}{2m}\,,\qquad E_n=\left(n+\frac{\hbar\lambda}{2m\omega}+\frac{1}{2}\right)\,\hbar\,\omega$$ where $H_n$ is the Hermite polynomial. The effect of the second term in the energy equation is to shift the energy of the ordinary oscillator by the value $\frac{\hbar^2\lambda}{2m}$. The solution of Eq.(12) is
\begin{equation}
\chi(t)=C\,e^{\frac{m\omega}{2\hbar}\,\, c^2t^2}\,,
\end{equation}
where $C$ is constant, and $\lambda=-\frac{m\omega}{\hbar}\,.$

The full wavefunction in 1+1 dimension for a  harmonic oscillator is thus
\begin{equation}
\psi_{0n}(x, t)=C\,\left(\frac{\sqrt{\frac{m\omega}{\pi\,\hbar}}}{2^nn!}\right)^{1/2}H_n\left(\sqrt{\frac{m\omega}{\hbar}}\,x\right)\,e^{-\frac{m\omega}{2\hbar}(x^2-c^2t^2)}\,,
\end{equation}
with a  total energy $E_*=n\,\hbar \,\omega$.
\\
In 3+1 dimensions, the total energy of the oscillator is given by
\begin{equation}
E_*= \left(n+\frac{3}{2}\right)\,\hbar\,\omega-\frac{1}{2}\hbar\,\omega\qquad {\rm or} \qquad\qquad E_{*n}=(n+1)\,\hbar\,\omega\,.
\end{equation}
It is intriguing to state that the total energy of the harmonic oscillator in 3-dimensions  was considered to emerge from the spatial coordinates and the corresponding momenta contribution. It was never argued that, as in special relativity, time is also a coordinate that should be incorporated. Hence, a relativistic harmonic oscillator (4-dimensional) should have its energy being contributed from the 4 coordinates. And since time is considered as an imaginary coordinate, its contribution to the total energy is negative and equals to $-\frac{1}{2}\,\hbar\,\omega$. This assertion is tantamount to the \emph{equipartition theorem} which states that each squared coordinate (degree of freedom)  in the Hamiltonian of the system contributes a value of $\frac{1}{2}\, k_BT$ to the total energy of the thermal system, where $T$ is the absolute temperature and $k_B$ is the  Boltzman constant. We analogously state the following:

"\emph{for a relativistic  quantum harmonic oscillator each squared coordinate contributes an energy of $\frac{1}{2}\, \hbar\,\omega$ to the total energy, while a time coordinate contributes  $-\frac{1}{2}\,\hbar\,\omega$, and each squared momentum along a direction $i$ contributes a term $n_i\,\hbar\omega$ to the total energy of the oscillator}".

 It is found recently that a complex harmonic oscillator in 2-dimensions has an energy of $E_n=(n+1)\hbar\omega$, sharing the same energy as  that of a 3+1 dimensional harmonic oscillator \textcolor[rgb]{0.00,0.07,1.00}{\cite{arb}}.

\section{\textcolor[rgb]{0.00,0.07,1.00}{Relativistic quantum interval}}

In special relativity, the worldline of a particle in space-time is  determined by the squared interval, $s^2=c^2t^2-r^2$. This interval is invariant under Lorentz transformation. The worldline of a massless particle (photon) is determined by $s^2=0$. This means that the photon moves with velocity of light. Massive objects move along a worldline (time-like) that is causally connected, since it moves with a velocity less than the velocity of light in vacuum ($v<c$). However, in quantum mechanics, the particle worldline is a bit fuzzy, since the Heisenberg uncertainty relation, relating position and momentum, doesn't  render the particle path to be well-defined.

Assuming that Eq.(6) satisfies a plane wave solution of the form $\psi_0(r, t)=A\,e^{-i(\omega'\,t-\vec{k}\cdot\vec{r})\,}$, where $A$ is constant, we obtain
\begin{equation}
\omega^2(r^2-c^2t^2)\,m^2-2E\,m +\hbar^2(k^2-\frac{\omega'^2}{c^2})=0\,.
\end{equation}
For massless oscillator (photon), Eq.(18) yields the non-dispersive relation, $\omega'=k\,c$. Equation (18) can be seen to define the mass of the oscillator. And since the mass of the oscillator is  positive definite, then for a minimum mass, one finds
\begin{equation}
c^2t^2-r^2=\lambda^2_c\left(1-\frac{k^2c^2}{\omega^2}\right)\,,\qquad\qquad \lambda_c=\frac{\hbar}{mc}\,,\qquad \omega'=\omega\,\,.
\end{equation}
Equation (19) defines the squared quantum interval of a massive oscillator. It is time-like for $\omega<kc$ and time-like for $\omega>kc$\,, and light-like (null) when $\omega=ck$\,. Equation (19) can be related to the refractive index ($n_r$), where $n_r=kc/\omega$, as
\begin{equation}
c^2t^2-r^2=\lambda^2_c\left(1-n_r^2\right)\,.
\end{equation}
This equation blends geometry, relativity, optic and quantum mechanics.
It is argued by Korbel  that the above interval expresses the particle space-time extensions \textcolor[rgb]{0.00,0.07,1.00}{\cite{comp}}. Furthermore, this space-time is related to the internal structure of  the oscillator. Owing to Eq.(19), his model is applicable for stationary oscillator, \emph{i.e.,} that with $k=0$\,, where $\lambda_c=s$.
Hence, the interval squared depends on the medium in which the oscillator travels. We prefer to call Eq.(20) the quantum interval. It reveals that the space-time is not flat inside a medium, and will be curved if the oscillator (\emph{e.g.,} light) had a mass. Equation (20) represents an equation of a hyperbola, or straight line depending on the value of $n_r$.

The interval is space-like ($s^2<0$) for $n_r>1$, and time-like ($s^2> 0$) when $n_r<1$.  Consequently, the mass of a particle (oscillator) reflects its interaction with the medium in which it moves. In a medium with $n_r<1$, the role of space and time is interchanged. This situation occurs for a particle falling inside a black hole beyond the event horizon. This is tantamount to saying that time-like interval becomes space-like and vice versa. Interestingly, conductors have $n_r<1$ for x-rays and ultra-violet (UV) radiations. This means that conductors are opaque to ordinary light (oscillator), but transparent to x-rays and UV radiations.

\subsection{\textcolor[rgb]{0.00,0.07,1.00}{Mass spectrum of oscillator}}
Let us now apply Eq.(17) in Eq.(8), using Eq.(20), to obtain mass spectrum of the oscillator as
\begin{equation}
m_{*n}^2=m_\omega^2\left(n_r^2-1-\beta\,\left(n+1\right)\right)\,,\qquad m_\omega=\frac{\hbar\omega}{c^2}\,,\qquad \beta=\frac{2mc^2}{\hbar\,\omega}\,.
\end{equation}
One can suggest that $m$ is the mass of the oscillator in 3 dimensions whereas $m_*$ its mass in 4- dimensions. It is apparent from Eq.(21) that the mass of higher excitations  becomes smaller.  The parameter $\beta$  signifies the number of particle - antiparticle pair created at a given oscillation ($\omega$), and $m_\omega$ is the equivalent mass to a given oscillation. It is emphasised by Aldaya \emph{et al}. that the quantum dynamics of a relativistic oscillator involves an extra quantization condition that $N=\frac{mc^2}{\hbar\,\omega}=1\,, \frac{3}{2}\,, 2\,, \cdots$ \textcolor[rgb]{0.00,0.07,1.00}{\cite{relativ5}}. Since the minimum energy to create a single pair requires an energy of $2mc^2=\hbar\,\omega$\,, the minimum vale of $\beta$ is 1. Interestingly, Eq.(21) represents the quantization of the mass of the oscillator in 3+1 dimensions \textcolor[rgb]{0.00,0.07,1.00}{\cite{mass}}. This mass quantization may reflect the mass spectrum of a bound system that is composed of harmonic oscillators (\emph{e.g.,} hadrons) \textcolor[rgb]{0.00,0.07,1.00}{\cite{relativ3}}. For a bound system, the interval $s^2=\ell^2$ in Eq.(20), can express the physical size (extension) of the oscillator, $\ell=\lambda_c\sqrt{1-n_r^2}\,.$
For time-like interval $\ell$ is real, and will be imaginary for space-like interval.
Thus, the potential energy in Eq.(10) can be written as
\begin{equation}
V(r, t)=\frac{(n_r^2-1)}{\beta}\,\hbar\,\omega\,,\qquad\qquad V(r,t)=\frac{1}{2}\,m(n_r^2-1)\lambda_c^2\omega^2\,,
\end{equation}
upon using Eq.(21). It can be seen as representing an oscillation with an amplitude, $A=\sqrt{n_r^2-1}\,\lambda_c$. It is evident that Eq.(22) reveals the quantum nature of the oscillator potential energy.  A negative potential energy would suggest that the force between the oscillators becomes repulsive. A quantum force of the form $V/\ell\propto \hbar\,\omega$ can be associated with these oscillators. This may explain why no free quark can be observed, and that the quarks (if modelled by oscillators) are confined perpetually inside hadrons. The quantum potential energy in Eq.(22) can be compared with that due to spin-orbit coupling, or the one due to a magnetic dipole moment in presence of an external magnetic field ($B$) \textcolor[rgb]{0.00,0.07,1.00}{\cite{dirac-klein}}.

The  motion  of  a  charged  particle ($q$)  in  a  uniform  magnetic  field was considered by Landau who showed that  the  motion  parallel  to  the  field  is  unaffected,  while that perpendicular to the field  is  that  of  a  two-dimensional  harmonic  oscillator, the  frequency  of  which  is  the   cyclotron  frequency \textcolor[rgb]{0.00,0.07,1.00}{\cite{landau, landau1}}. This is given by $\omega=qB/m$, so that Eq.(22) becomes
\begin{equation}
V=-\mu\,\,B\,,\qquad\qquad \mu=(1-n_r^2)\frac{q\hbar}{m\beta}.
\end{equation}
In this situation $\mu$ defines the magnetic dipole moment of the oscillator, which depends on the medium in which the oscillator pair exists.
It seems that the effect of space and time on a harmonic oscillator is analogous to the influence of the electric and magnetic fields, respectively. A negative potential energy arises when $n_r<1$. This indicates that the emerging force is attractive. The oscillator tends to be frequency under low vibration frequency. One may associate a magnetic moment of a stationary oscillator, \emph{i.e.}, $k=0$ and $n_r=0$\,. For $\beta=1$, or $\hbar\,\omega=2mc^2$, one has $\mu_s=\frac{q\hbar}{m}$, and for  $\beta=2$, or $\hbar\,\omega=mc^2$, one has $\mu_l=\frac{q\hbar}{2m}$\,. The former one can be related to  spin and the latter to orbital angular momentum of the oscillator.

\section{\textcolor[rgb]{0.00,0.07,1.00}{Concluding Remarks}}

We  have considered in this work a model of quaternionic  quantum harmonic oscillator using Shr\"odinger paradigm. This model can be applied to study the prorogation of massive photons in space-time. We obtained a relativistic quantum harmonic oscillator (in 3+1 dimension) generalizing the non-relativistic Shr\"odinger equation. The total energy of this oscillator is given by $E_{*n}=(n+1)\,\hbar\,\omega$. It can be seen as a time-correction to the non-relativistic oscillator, where the correction energy is $-\frac{1}{2}\,\hbar\,\omega$.

It is shown that a massless photon follows a null interval, and thus travels in straight line. However, a massive photon bends follows a time-like interval in a medium, and therefore bends. Hence, inheritantly all (waves) fields curve inside a medium whether massless or massive. We have found that the squared interval of a harmonic oscillator is related to medium properties in a quantum mechanical manner. Moreover, the ground state energy of the harmonic oscillator in 1+1 dimensions vanishes. The mass of the oscillator in 3+1 dimensions is found to be quantized.

\section{Acknowlegements}

I would like to thank Dr. I. B. Tomsah for reading the manuscript and interesting discussions.

\end{document}